\documentclass[journal=jacsat,manuscript=article]{achemso}

\usepackage{chemformula} %
\usepackage[T1]{fontenc} %
\usepackage[utf8]{inputenc}
\usepackage{textcomp}
\usepackage{amsmath,amssymb}

\DeclareUnicodeCharacter{2005}{}

\author{Jenna C. Fromer}
\author{Connor W. Coley}
\email{ccoley@mit.edu}
\affiliation[Unknown University]
{Department of Chemical Engineering, MIT, Cambridge, MA 02139}
\alsoaffiliation[Second University]
{Department of Electrical Engineering and Computer Science, MIT, Cambridge, MA
02139}

\title[]
  {Computer-Aided Multi-Objective Optimization in Small Molecule Discovery}

\abbreviations{}

\begin{document}

% \section{Bigger Picture}
% \input{bigger-picture}

\begin{abstract}

Molecular discovery is a multi-objective optimization problem that requires identifying a molecule or set of molecules that balance multiple, often competing, properties. %
Multi-objective molecular design  is commonly addressed by combining properties of interest into a single objective function using scalarization, which imposes assumptions about relative importance and uncovers little about the trade-offs between objectives. In contrast to scalarization, Pareto optimization does not require knowledge of relative importance and reveals the trade-offs between objectives. However, it introduces additional considerations in algorithm design. %
In this review, %
we describe pool-based and \textit{de novo} generative approaches to  multi-objective molecular discovery with a focus on Pareto optimization algorithms. We show how pool-based molecular discovery is a relatively direct extension of multi-objective Bayesian optimization and how the plethora of different generative models extend from single-objective to multi-objective optimization in similar ways using non-dominated sorting in the reward function (reinforcement learning) or to select molecules for retraining (distribution learning) or propagation (genetic algorithms). %
Finally, we discuss some remaining challenges and opportunities in the field, emphasizing the opportunity to adopt Bayesian optimization techniques into multi-objective \textit{de novo} design. %
\end{abstract} %

  \section{Introduction}
\begin{sloppypar}

Molecular discovery is inherently a constrained multi-objective optimization problem. Almost every molecular design application requires multiple properties to be optimized or constrained. For example, for a new drug to be successful, it must simultaneously be potent, bioavailable, safe, and synthesizable. Multi-objective optimization, also referred to as multi-parameter optimization (MPO), pertains to other applications as well, including solvent design \cite{chong_design_2022, ten_computer-aided_2021, papadopoulos_multiobjective_2006, mah_design_2019}, personal care products \cite{yee_optimization_2022, ooi_design_2022}, electronic materials \cite{karasuyama_computational_2020, devereux_chapter_2021,  hautier_finding_2019, hachmann_harvard_2011, ling_high-dimensional_2017}, functional polymers \cite{jablonka_bias_2021, mannodi-kanakkithodi_multi-objective_2016}, and other materials \cite{hanaoka_bayesian_2021, solomou_multi-objective_2018, khatamsaz_multi-objective_2022}. 
Redox-active species in redox flow batteries must maximize redox potential and solubility to ensure a high cell voltage \cite{kowalski_recent_2016, winsberg_redoxflow_2017}. 
Sustainability of new materials (e.g., emissions caused during production and disposal \cite{fleitmann_cosmo-suscampd_2021}) is also an increasingly important design objective \cite{wilson_accelerating_2022, melia_materials_2021}, which is particularly important for working fluids \cite{raabe_molecular_2019, kazakov_computational_2012,fleitmann_cosmo-suscampd_2021}. Multi-objective optimization can address multiple design criteria simultaneously, allowing for the discovery of molecules that are most fit for a specific application. 

When many objectives must be optimized simultaneously, a common approach is to aggregate the objectives into a single objective function,  which requires quantifying the relative importance of each objective. This method, also known as \emph{scalarization}, reduces a multi-objective molecular optimization problem into one that is solvable with single-objective algorithms, but the ability to explore trade-offs between objectives is limited. Further, the optimization procedure must be repeated each time the scalarization function is adjusted. In contrast, Pareto optimization, which discovers a set of solutions that reveal the trade-offs between objectives, relies on no prior measure of the importance of competing objectives. This approach allows an expert to modify the relative importance of objectives without sacrificing optimization performance or repeating the optimization procedure. The solution set of a Pareto optimization contains the solution to every scalarization problem with any choice of weighting factors. For these reasons, we believe that Pareto optimization is the most robust approach to multi-objective molecular discovery. 

The discovery of optimal molecules can be framed as either a search for molecules from an enumerated library or generation of novel molecules (i.e., \textit{de novo} design) \cite{sridharan_modern_2022, meyers_novo_2021}. The extension of both discovery approaches from single-objective to multi-objective optimization has been reviewed for molecular discovery \cite{segall_multi-parameter_2012, nicolaou_molecular_2007} and more specifically drug discovery \cite{ekins_evolving_2010, nicolaou_multi-objective_2013}. However, recent developments, specifically in \textit{de novo} design using deep learning, warrant further discussion and organization of new methods. 

In this review, we organize established and emerging multi-objective molecular optimization (MMO) techniques. %
After defining MMO and introducing relevant mathematical concepts, we describe key design choices during the formulation of an optimization scheme. Then, we provide a thorough discussion of relevant methods and case studies, first in library-based optimization and then in \textit{de novo} design. %
Finally, we share some open challenges in MMO and propose future work that we believe would most advance the field.

\end{sloppypar}

\section{Defining Multi-Objective Molecular Optimization}

The molecular discovery literature is riddled with approaches to solve the inverse problem of property $\rightarrow$ structure, many of which are labeled ``multi-objective''. However, the line between  multi-objective molecular optimization (MMO) and single-objective or constrained optimization is quite blurred. To organize the field's communication of MMO methodologies, we classify MMO as follows: %
\begin{enumerate}
    \item Multiple objectives, which are not aggregated into a single scalar objective, are considered. Some trade-off exists between objectives (i.e., they are not perfectly correlated). 
    \item The domain over which to optimize (``design space'') is a chemical space. Molecules in this space may be defined either implicitly (e.g., as latent variables that can be decoded using generative models) or explicitly (i.e., as a molecular library).
    \item The goal of the optimization task is to identify molecules that maximize or minimize some molecular properties. We consider tasks that aim to identify molecules with properties within some specified range to be constrained generation, not multi-objective optimization.

\end{enumerate}
Any definitive scope of MMO is bound to be somewhat subjective. Yet, we believe the preceding definition captures all relevant implementations of MMO and excludes methods that are better categorized elsewhere (e.g., as a single-objective optimization or constrained optimization). 

Exhaustive screening for multiple optimized properties, typically referred to as virtual screening \cite{rizzuti_chapter_2020}, can be viewed as an inefficient approach to MMO. This approach has been used to identify multi-target inhibitors \cite{wei_multiple-objective_2019, ramsay_perspective_2018, kim_two-track_2022} as well as selective inhibitors \cite{kuck_novel_2010}. In the interest of summarizing efficient optimization algorithms, we do not discuss enumeration and exhaustive screening approaches in this review.

\section{Preliminary Mathematical Concepts in MMO}
    \subsection{The Pareto front}
\begin{figure}
        \centering
        \includegraphics[trim={0cm 0 0cm 0},clip]{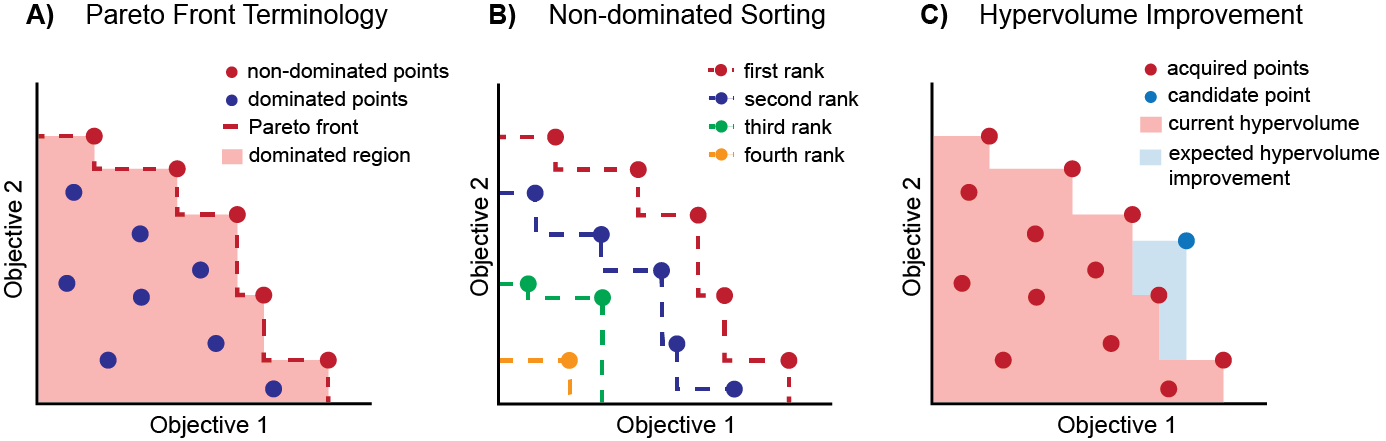}
        \caption{Terminology and acquisition functions in pareto optimization. (A) Visual depiction of common Pareto terminology including the Pareto front, dominated and non-dominated points, and dominated region. The area of the dominated region is the hypervolume. (B) Non-dominated sorting, also referred to as Pareto ranking. (C) Hypervolume improvement for one candidate point over the current hypervolume defined by the set of previously acquired points in the absence of uncertainty. }
        \label{fig:basic_pf_acq}
    \end{figure}

    In MMO problems, two or more desirable molecular properties compete with one another. For \emph{Pareto optimal} solutions, an improvement in one objective is detrimental to at least one other objective. For instance, when a selective drug is designed, strong affinity to the target and weak affinity to off-targets are both desired. However, when the binding affinities to on- and off-targets are highly correlated (i.e., they bind strongly to similar molecules), an increase in potency to the target often necessitates a decrease in selectivity. The \emph{Pareto front} quantifies (and, in the  2- or 3-objective case, visualizes) these types of trade-offs. Figure~\ref{fig:basic_pf_acq}A illustrates a Pareto front for two objectives which are to be maximized, with points in red representing the \emph{non-dominated points}, which form the Pareto front and define the set of optimal solutions for the multi-objective optimization problem. For these points, an improvement in one objective necessitates a detriment to the other objective. One can imagine that each objective is a desired property and that each point on the plot represents one molecule. For simplicity and ease of visualization, we always consider that objectives are maximized for the remainder of the review. Pareto fronts for minimized objectives would instead appear in the lower left corner, as opposed to the upper right.

    The \emph{hypervolume} of a set is the volume spanned by the Pareto front with respect to a reference point. In the 2-dimensional case, the hypervolume is the area that is dominated by the Pareto front (the red shaded region in Figure~\ref{fig:basic_pf_acq}AC). This metric can evaluate how ``good'' a Pareto front is: a larger hypervolume indicates a larger dominated region (i.e., a ``better'' Pareto front). 
    
\begin{figure}
        \centering
        \includegraphics[scale=1.5]{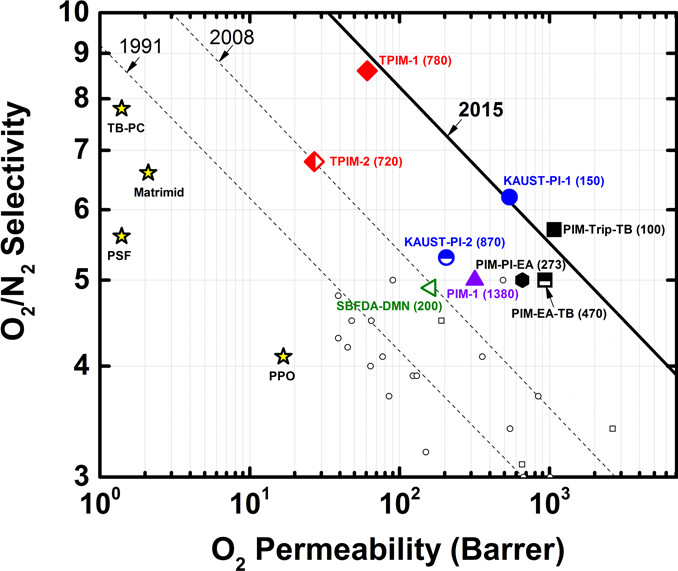}
        \caption{Progress in membranes for gas separation as revealed by the movement of a Pareto front. Reproduced from \citet{swaidan_fine-tuned_2015}.}
        \label{fig:membrane_PF}
    \end{figure}

    Progress in new materials development is often reported and visualized by the advancement of a Pareto front. As an example, in gas separation applications, membrane selectivity and permeability are two competing objectives which are both to be maximized. The trade-offs for this optimization can be visualized as a Pareto front. Figure~\ref{fig:membrane_PF} shows the improving upper bound for the two maximized objectives, which can be understood as an expansion of the Pareto front from 1991 to 2015 \cite{swaidan_fine-tuned_2015}. 
    
    \subsection{Single-objective Bayesian optimization}  
\begin{figure}
        \centering
        \includegraphics[scale=1, trim={0cm 0cm 0cm 0cm},clip ]{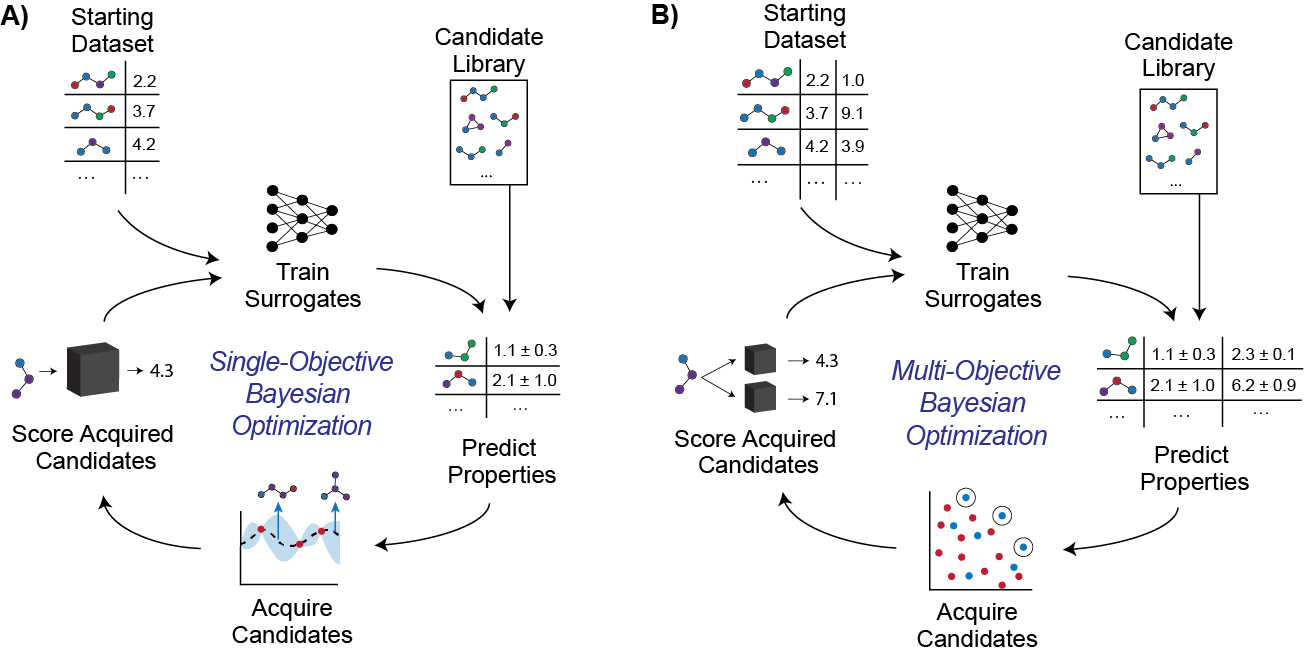}
        \caption{Overview of the Bayesian Optimization workflow and the commonalities between the (A) single-objective and (B) multi-objective settings.}
        \label{fig:BO}
    \end{figure}

Bayesian optimization (BO) is a strategy for black box optimization where the scalar function to be optimized, sometimes referred to as the \emph{oracle}, may be non-differentiable or difficult to measure (costly) \cite{frazier_bayesian_2018}. The workflow of Bayesian optimization applied to single-objective molecular discovery is summarized in Figure~\ref{fig:BO}A. %
    
    BO is an iterative optimization procedure that begins by defining some prior model to map the design space to the objective. This model is called a \textit{surrogate model} and, in the molecular setting, is equivalent to a quantitative %
    structure-property relationship (QSPR) model.
    The surrogate model is used to predict the objective values of hypothetical candidates in the design space, which an \textit{acquisition function} uses (along with the surrogate model uncertainty) to prioritize which candidates %
    to sample next. The newly sampled, or \emph{acquired}, molecules are then evaluated, or \emph{scored}, against the oracle, and this new data is used to refine the surrogate model. The process is repeated until some stopping criterion is met: the objective value of the acquired molecules converges, resources are expended, or some objective value threshold is attained.

    The acquisition function is central to BO. This function quantifies the ``utility'' of performing a given experiment and can be broadly understood to balance both the \emph{exploitation} and \emph{exploration} of the design space \cite{shahriari_taking_2016}.
    In molecular BO, exploration prevents stagnation in local optima and can encourage acquisition of more diverse molecules.
   However, the acquisition function must also exploit, selecting candidates predicted to optimize the objective, which enables the algorithm to converge upon an optimum and identify the best-performing molecules. A few acquisition functions for the case where a single objective ($f$) is maximized are worth mentioning: 
    \begin{enumerate}
        \item Expected improvement (EI): \begin{equation}
            \text{EI}(x) = \mathbb{E}[\max\{0,f(x)-f^*\}],  \end{equation}
            in which $f(x)$ represents the objective value for some molecule $x$, $\mathbb{E}$ is the expectation operator, and $f^*$ is the best objective value attained so far from the acquired molecules \cite{frazier_bayesian_2018, shahriari_taking_2016}. 
        \item Probability of improvement (PI): 
            \begin{equation} \text{PI}(x) = \mathbb{E}[(f(x)-f^*)>0] \end{equation} The PI metric estimates how likely a new molecule $x$ is to outperform the current best molecule \cite{shahriari_taking_2016}. 
        \item Greedy acquisition (G):
            \begin{equation} \text{G}(x) = \hat f(x) \end{equation} Here, the acquisition function is simply the predicted value for the objective function, regardless of uncertainty and what has been observed so far \cite{pyzer-knapp_bayesian_2018}. 
        \item Upper confidence bound (UCB): 
            \begin{equation}
            \text{UCB}(x) = \hat f(x) + \beta \sigma(x),  \end{equation}
            in which $\sigma$ is the surrogate model prediction uncertainty and $\beta$ is a hyperparameter \cite{shahriari_taking_2016}. 
    \end{enumerate}
    
    \begin{sloppypar}
    While the BO literature thoroughly discusses and tests many acquisition functions, we have only described a few which are most popular in MMO. We refer readers interested in single-objective acquisition functions to \citeauthor{frazier_bayesian_2018}'s tutorial \cite{frazier_bayesian_2018} or \citeauthor{shahriari_taking_2016}'s review \cite{shahriari_taking_2016}. 
          \end{sloppypar}

    \subsection{Multi-objective Bayesian optimization}

    Pareto optimization problems, in which multiple objectives are considered simultaneously without quantification of relative objective importance, must be handled with a slightly modified set of tools, although the core BO ideology remains the same (Figure~\ref{fig:BO}B). First, all oracle functions must be approximated either with multiple surrogate models, a multi-task surrogate model \cite{shahriari_taking_2016}, or some combination thereof.
    Second, the acquisition function must account for all objectives without explicitly assigning a relative importance weight to each of them. Here, the goal is to
    expand the Pareto front, or increase the dominated hypervolume, as much as possible. We focus on three multi-objective acquisition functions: 
\begin{enumerate}
    \item Expected hypervolume improvement (EHI):  
     \begin{equation}
         \text{EHI}(x) = \mathbb{E}[\max(0, \text{HV}(\mathcal{X}_{acq} \cup \{x\} )-\text{HV}(\mathcal{X}_{acq}))],
     \end{equation}
     in which HV is the hypervolume and $\mathcal{X}_{acq}$ is the set of previously acquired candidates. EHI is best understood as an analog to the single-objective expected improvement which measures improvement in hypervolume instead of objective value.  
    
    \item Probability of hypervolume improvement (PHI): 
         \begin{equation}
         \text{PHI}(x) = \mathbb{E}[(\text{HV}(\mathcal{X}_{acq}\cup \{x\})-\text{HV}(\mathcal{X}_{acq})) > 0]
     \end{equation}
    PHI, comparable to probability of improvement, is the probability that an acquired point will improve the hypervolume by any amount.

    \item Non-dominated sorting (NDS): NDS assigns an integer rank to each molecule by sorting the set of molecules into separate fronts. One can imagine identifying a Pareto front from a finite set of molecules (denoted first rank), removing that Pareto front, and subsequently identifying the next Pareto front (denoted second rank), as shown in Figure~\ref{fig:basic_pf_acq}B. The assigned Pareto rank to each molecule is taken to be its acquisition score. %
    NDS does not consider uncertainty, and a candidate's assigned Pareto rank is taken to be its acquisition score. The first rank candidates are equivalent to the set of points that would be acquired from using greedy acquisition with every set of possible scalarization weights, so NDS can be thought of as a multi-objective analog of greedy acquisition.

\end{enumerate}

    \subsection{Batching and batch diversity}
    
    While the canonical BO procedure evaluates candidates sequentially by acquiring the single candidate with the highest acquisition score at each iteration, many molecular oracles can be evaluated in batches. %
    Experiments performed in well plates are naturally run in  parallel, and expensive computations are often distributed in batches to make the best use of computational resources. In the BO workflow, this means that an acquisition function should be used to select \emph{a set} of molecules, instead of just one. A naïve approach, \emph{top-$k$} batching, scores molecules normally and acquires the $k$ candidates with the highest acquisition scores. The utility of the entire set is thus implicitly taken to be the sum of individual acquisition scores. However, the information gained from acquiring one molecule that is highly similar to another molecule in the batch is likely to be small. 
    
    In batched multi-objective optimization, the acquisition function should maximize the utility of scoring the \emph{entire batch}. %
    For the case of acquisition with EHI, this refers to the improvement in hypervolume after \emph{all} molecules in a batch are acquired. %
    One can imagine that acquiring a set of candidates very near each other on the Pareto front would not maximize this utility. An ideal batching algorithm would consider all possible batches, predict the utility of each, and select the batch with greatest utility. However, solving this combinatorial optimization exactly is intractable. Instead, approximations are used to construct batches iteratively: identify the most promising molecule, assume it has been observed, select the next most promising molecule, and repeat this until the desired batch size is achieved \cite{hiot_kriging_2010}. 
    
\begin{figure}[ht]
        \centering
        \includegraphics[scale=1]{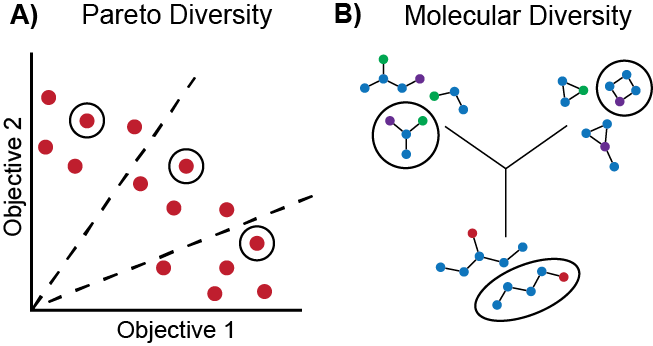}
        \caption{Comparing (A) Pareto diversity and (B) molecular/structural diversity for batch acquisition. Promoting one form of diversity does not necessarily improve the other.}
        \label{fig:diversity}
    \end{figure}

    Batched optimization is more often approached with heuristics that promote some measure of \emph{diversity} within a batch while selecting molecules with high acquisition scores. For example, the objective space can be split into regions (Figure~\ref{fig:diversity}A) with a limit on the number of candidates acquired in each region \cite{konakovic_lukovic_diversity-guided_2020, deb_evolutionary_2014}; likewise, candidates in less crowded regions along the Pareto front can be more strongly favored \cite{deb_fast_2002}. Such approaches to promote \emph{Pareto diversity} have been incorporated into multi-objective molecular design  \cite{verhellen_graph-based_2022, agarwal_discovery_2021, grantham_deep_2022}.
    
    Diversity of the \emph{design space} can also be considered during acquisition, which is distinct from Pareto diversity and can also be applied to single-objective optimization \cite{gonzalez_new_2022}. 
    In MMO, design space diversity is equivalent to the the \emph{structural}, or \emph{molecular}, diversity of a batch (Figure~\ref{fig:diversity}B). %
    Molecular diversity can be measured with metrics like Tanimoto similarity using fingerprint representations, which characterize a specific kind of structural similarity. As with Pareto diversity, structural diversity constraints can be imposed during acquisition \cite{janet_accurate_2020, nicolaou_novo_2009}.    
    While one might predict that Pareto front diversity also indicates molecular diversity, this is not necessarily true. It is possible for two structurally similar molecules to have different properties and therefore lie in different regions of the objective space; conversely, molecules with similar properties are not necessarily structurally similar. %

\section{Formulating Molecular Optimization Problems}

\begin{figure}
    \centering
    \includegraphics[scale=1, trim={0cm 0cm 0cm 0cm},clip ]{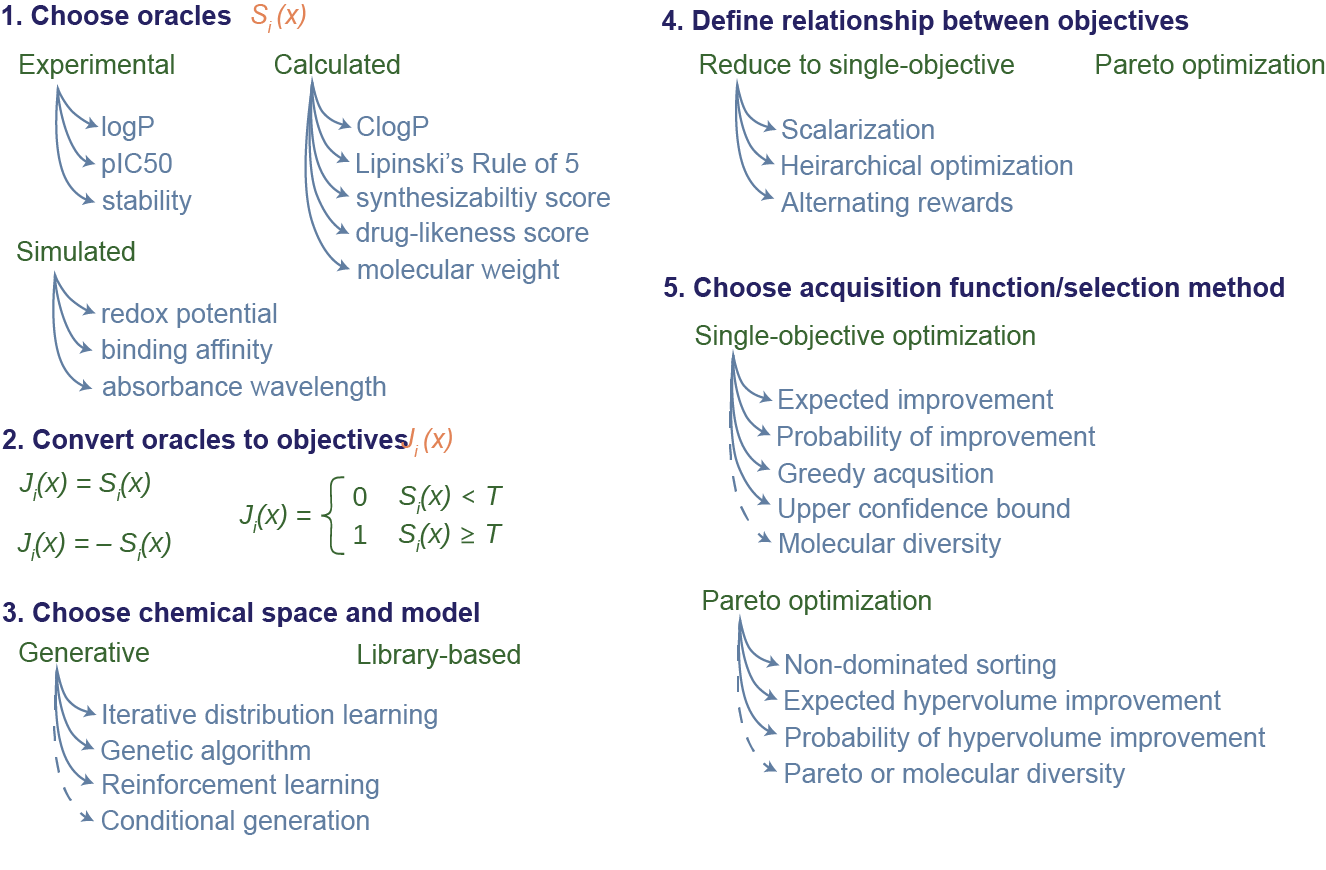}
    \caption{Decisions when formulating MMO problems. As discussed further in later sections, iterative generative models employ selection criteria for retraining or propagation, which are analogous to acquisition functions in Bayesian optimization. Conditional generation, although capable of proposing molecules with a specified property profile, is non-iterative and therefore does not utilize selection criteria or an acquisition function. Single-objective acquisition functions can only consider molecular diversity, while Pareto acquisition functions can consider both molecular and Pareto diversity.}
    \label{fig:formulating}
\end{figure}

A molecular optimization task always begins with some statement of desired properties. Some of the subsequent formulation decisions are listed in Figure~\ref{fig:formulating}. First, the individual properties must be converted to mathematical objectives. Then, the means of proposing candidate molecules, either \textit{de novo} or library-based, must be selected. If more than one objective exists, they must either be aggregated into a single objective or treated with an appropriate multi-objective formulation. Finally, an acquisition function, or selection criterion in the case of \emph{de novo} design, must be selected. In this section, we explore some of these design choices in detail. 

    \subsection{Converting a desired property to a mathematical objective function}
    In the formulation of any MMO task, after properties of interest are identified by a subject matter expert, the individual objectives must be quantitatively defined (Figure~\ref{fig:formulating}, Panel 2). While this seems like an easy task, framing the objectives can be subjective in nature. 
    If one property of interest for a molecular optimization task is estimated by a score $S(x)$, there are still multiple ways to represent the corresponding value to be maximized ($J(x)$), including but not limited to: 
    \begin{enumerate}
        \item A continuous, strictly monotonic treatment, where a greater value is strictly better:
        \begin{equation}
            J(x) = S(x)
        \end{equation}
        \item A thresholded, monotonic  treatment, where some minimum $T$ is required:
            \begin{equation}
                J(x) =  \left\{
                \begin{array}{ll}
                    S(x) & \quad S(x) \geq T \\
                    -\infty & \quad S(x) \leq T
                \end{array}
            \right.
            \end{equation}
        
        \item A Boolean treatment, where some minimum $T$ is required and no preference is given to even higher values:            \begin{equation}
                J(x) =  \left\{
                \begin{array}{ll}
                    1 & \quad S(x) \geq T \\
                    0 & \quad S(x) \leq T
                \end{array}
            \right.
            \end{equation}
    \end{enumerate}
    
    The most appropriate representation depends on the property of interest and the application, demonstrated here for common properties of interest for novel drug molecules. If $S$ predicts a ligand's binding affinity to a target protein, a higher affinity is often better, so the first representation may be most appropriate. If $S$ predicts solubility, there may be no additional benefit of greater solubility once a certain solubility is met that allows for sufficient delivery and bioavailability. In this case, the third representation, which is most consistent with a property constraint instead of an optimized objective, would be most fitting. In a similar manner, remaining components of Lipinski's Rule of 5 \cite{lipinski_experimental_2001} define some threshold, and no extra benefit is attained once the threshold is met. These heuristics may be most appropriately defined as constraints and not optimized objectives.   %
    
    The perspectives of domain experts during objective formulation are extremely valuable to ensure that molecules identified as optimal are suitable for the application. However, in cases where expertise is not available or a specific threshold is unknown, we argue that solving the problem with a simple continuous representation (Representation 1) is most robust because it requires no predefined hyperparameters or assumptions. This way, constraints can later be imposed on the solution set without needing to repeat the optimization from scratch.

    \subsection{Choosing between library-based selection and \textit{de novo} design}
    
    Once the objectives are defined, an approach to chemical space exploration must be chosen. The scope of exploration can be limited to an explicitly defined molecular library, which can be constructed to bias exploration toward chemical spaces relevant to a specific task. Alternatively, a \emph{\textit{de novo}} design tool can be used to ideate novel molecules not previously seen or enumerated.     
    The type of generative model influences the area of chemical space that is explored \cite{coley_defining_2021}. %
    For example, the chemical space explored by genetic algorithms  %
    may be constrained by the molecules used as the initial population and the set of evolutionary operators that are applied to the population. In a more general sense, the molecules that can be generated by any \textit{de novo} model will be determined by the training set and many other design choices. %
    Care can be taken to ensure that the chemical space explored is sufficient for the given task. %
    
    \subsection{Defining the relationship between different objectives}%
    Once individual objective functions are defined and the chemical space approach is chosen, the next challenge is to decide how to consider all objectives simultaneously. 
    The most naive choice %
    is to simply combine the objective functions into one aggregated objective function, referred to as \emph{scalarization}. The scalarized objective function is most commonly a weighted sum of objectives \cite{gomez-bombarelli_automatic_2018, winter_grunifai_2020, fu_mimosa_2021, hartenfeller_concept_2008, s_v_multi-objective_2022, ooi_integration_2018, liu_data-driven_2022}, with weighting factors indicating the relative importance of different objectives. A weighted sum of multiple binding affinities has been used to identify multi-target as well as selective inhibitors \cite{winter_efficient_2019}. 
    Nonlinear scalarization approaches are also utilized in MMO problems \cite{urbina_megasyn_2022, firth_moarf_2015, hoffman_optimizing_2022}. For example, \citeauthor{gajo_multi-objective_2018} divide predicted drug activity by toxicity to yield a scalarized objective function \cite{gajo_multi-objective_2018}. The objective function can also be framed as a product of Booleans \cite{chen_helix-mo_2022}, each of which denotes whether a given threshold is met.  This scalarization approach has been utilized to identify multi-target kinase inhibitors \cite{jin_multi-objective_2020}. %
    Booleans can also be summed to define an objective function, commonly referred to as multi-property optimization \cite{barshatski_multi-property_2021}.
    As with the definition of individual objectives, the scalarization function must be justified by the use case. 
    There are alternatives to scalarization that also reduce a multi-objective optimization into one that can be solved with single-objective algorithms, such as defining a hierarchy of objective importance \cite{hase_chimera_2018} or using alternating rewards to maximize each objective in turn \cite{goel_molegular_2021, pereira_optimizing_2021}.
    
    However, the solution to a scalarized multi-objective problem is equivalent to just a single point out of the many non-dominated solutions that exist on the Pareto front. %
    Scalarization is overly simplistic and requires a user to quantify the relative importance of different objective. It therefore fails to inform a user about the trade-offs between objectives. Even when the relative importance of objectives is known or can be approximated a priori, scalarization is strictly less informative than Pareto optimization which identifies the full set of molecules that form a Pareto front. We focus exclusively on Pareto optimization approaches to molecular discovery throughout the remainder of this review.

\section{Examples of MMO from Virtual Libraries}

   Library-based multi-objective molecular optimization aims to identify the Pareto front (or a set close to the Pareto front) of a large molecular library while scoring few molecules with the objectives. The well-established Bayesian optimization workflow (Figure~\ref{fig:BO}B) is exemplified by the retrospective studies of \citet{del_rosario_assessing_2020} and \citet{gopakumar_multi-objective_2018}. In general, the iterative optimization scheme entails training a surrogate model to predict properties of interest, selecting molecules for acquisition using surrogate model predictions and uncertainties, scoring the acquired molecules with the ground-truth objectives, and retraining the surrogate model. %
    
    \citet{janet_accurate_2020} apply this methodology to discover transition metal complexes for redox flow battery applications with maximized solubility and redox potential. Ideal complexes must be soluble in polar organic solvents commonly used for flow batteries and have high redox potentials to yield sufficient cell voltage. The design space the authors explore is a combinatorial library of almost 3 million complexes. A neural network surrogate model predicts solubilities and redox potentials from feature vector representations of complexes \cite{janet_resolving_2017}. DFT calculations served as the oracle for both solubility and redox potential, and the expected hypervolume improvement acquisition function was used.  To encourage exploration of structurally diverse complexes, the top 10,000 performers according to EHI were clustered in feature space to identify and evaluate 100 medoids. Improvements of over three standard deviations from the initial random set of complexes were observed for both objectives in just five iterations, which the authors estimate to represent a 500x reduction in simulations compared to a random search. 
    
    In a similar vein, \citet{agarwal_discovery_2021} use library-based Pareto optimization to search for redox-active materials with minimized reduction potential and solvation free energy. A third objective penalized deviation from a target peak absorption wavelength of 375nm. Candidates were scored with expected hypervolume improvement, while crowding distance constraints ensured acquisition of a diverse set along the Pareto front. When retrospectively applied to a dataset of 1400 molecules, a random search required 15 times more evaluations than did Bayesian optimization to acquire molecules dominating 99\% of the total possible hypervolume. Then, a prospective search was performed on a set of 1 million molecules, with the prior dataset serving as the first set of acquired molecules. Of the 100 molecules acquired during prospective BO iterations, 16 new Pareto-optimal molecules were identified.

    Most pool-based MMO problems follow this exact workflow with minor variability in the choice of acquisition function and consideration of diversity. This approach works effectively and is almost guaranteed to outperform random search baselines. While there is certainly room for algorithmic improvement (e.g., increasing sample efficiency of surrogate models, exploring the effects of batch size and diversity), we expect that future work will largely focus on additional applications incorporating more meaningful objective functions and experimental validation.

\section{Examples of MMO using Generative Models}

    The primary drawback of pool-based MMO is the explicit constraint on the chemical space that can be accessed. \textit{De novo} design relaxes this constraint and can, in principle, explore a wider (and in some cases, arguably infinite) region of chemical space. In many generative models, molecules are proposed as SMILES/SELFIES strings, graphs, or synthetic pathways. Some generate novel molecules by decoding continuous embeddings into discrete molecular structures while others modify those already identified with discrete actions. %
    We focus not on the details of each model, but instead on how certain categories of models aid in the molecular optimization task. A reader interested in a detailed discussion of generative models, which is outside the scope of this review, is directed to other publications \cite{bilodeau_generative_2022, sanchez-lengeling_inverse_2018, alshehri_deep_2020, mouchlis_advances_2021}.  %

   The myriad of multi-objective \textit{de novo} design approaches noticeably lack standardization. Unlike library-based discovery where multi-objective optimization is a modest extension of Bayesian optimization, the adaptation of generative models to MMO is not nearly as straightforward. We therefore introduce another categorization scheme for case studies in this section.

    \begin{figure}
        \centering
        \includegraphics[scale=1]{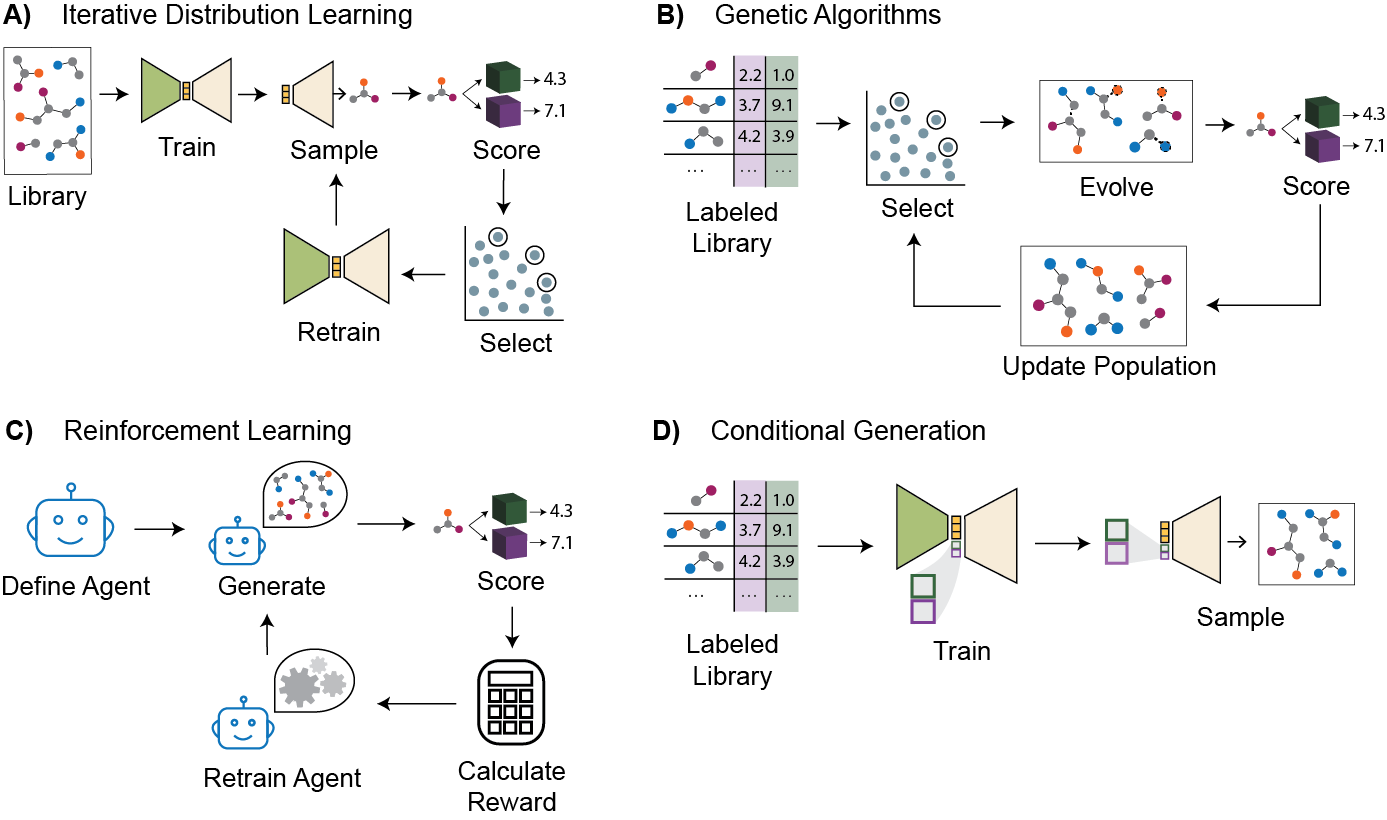}
        \caption{Optimization workflows for various generative model categories. Note that all model classes, except conditional generation, involve a scoring step and are designed to be iterative. The reward calculation step in reinforcement learning and the selection step in distribution learning and genetic algorithms are analogous to an acquisition function in multi-objective Bayesian optimization. While the termination criterion is not explicitly shown for distribution learning, genetic algorithms, and reinforcement learning, these iterative loops can accommodate various stopping criteria. We also emphasize that while an autoencoder architecture is depicted in both distribution learning and conditional generation, these generators can also be recurrent neural networks or other generative architectures.}
        \label{fig:generative}
    \end{figure}
    
    \subsection{Iterative retraining for distribution learning} Generative models that are designed for \emph{distribution learning} are intended to ideate molecules exhibiting a distribution of structures similar to those of the training set \cite{flam-shepherd_language_2022}. A very basic approach to optimization with an unsupervised generative model is to sample a set of molecules, evaluate their properties, and identify those that optimize the objective function; to extend this to multi-objective optimization, the Pareto front of the sampled set can be identified by evaluating all oracles \cite{frey_fastflows_2022}. This approach essentially uses a generative model to define a virtual library suitable for exhaustive screening. Optimization schemes can use distribution learning \emph{iteratively} to progressively shift the distribution of generated molecules and push the Pareto front. %
    To achieve this, generative models are iteratively retrained on the increasingly promising (e.g., closest to the Pareto front) subsets of the molecules they propose. %
    This process is akin to a simulated design-make-test loop, in which \emph{design} is analogous to sampling, \emph{make} to decoding to a molecule, and \emph{test} to evaluating the oracles.   %
    
    The iterative distribution learning workflow for single-objective optimization is exemplified by the library generation strategy defined by \citet{segler_generating_2018} to identify inhibitors predicted to be active against the 5-HT$_{2\text{A}}$ receptor. Here, a subset of molecules from the ChEMBL database, with corresponding experimental pIC$_{50}$ values against 5-HT$_{2\text{A}}$, was used to train both a SMILES-based recurrent neural network and a QSAR classifier to predict whether a molecule inhibits 5-HT$_{2\text{A}}$. Then, sequences of characters were randomly sampled from the RNN to generate SMILES representations of novel molecules. Molecules predicted by the QSAR classifier to be active were used to retrain the model, progressively biasing the generator to propose active molecules. After four iterations of retraining, 50\% of sampled molecules were predicted to be active, a significant increase from only 2\% in the initial random library. The same procedure has also been employed using a variational autoencoder to generate molecules with high docking scores to the DRD3 receptor\cite{boitreaud_optimol_2020}. 
    
    The extension of the method to multiple objectives is best illustrated by \citet{yasonik_multiobjective_2020} for the generation of drug-like molecules. As before, a recurrent neural network was pretrained to generate valid molecular SMILES strings. 
    Five oracles associated with drug-likeness were then minimized: ClogP (estimated lipophilicity), molecular weight, number of hydrogen bond acceptors, number of hydrogen bond donors, and number of rotatable bonds. A set of about 10k novel, unique, and valid molecules were sampled and scored according to the five properties. Non-dominated sorting was used to select half of these molecules for retraining. The use of NDS distinguishes this Pareto optimization from \citeauthor{segler_generating_2018}'s single-objective optimization. %
    Although continuous objective values were used during selection of  molecules for retraining
    , constraints associated with the oracles, derived from the ``Rule of Three'' \cite{congreve_rule_2003} (an extension of Lipinski's Rule of 5\cite{lipinski_experimental_2001}), were used to evaluate the generator's performance. After five retraining iterations, the fraction of molecules that fulfilled all five constraints increased from 2\% to 33\%. While there is no evidence that the Pareto front was shifted outwards (i.e., that the dominated hypervolume increased) after retraining iterations, this study demonstrates that a generative model's property distributions for multiple objectives can be shifted simultaneously. 
    
\begin{figure}
        \centering
        \includegraphics[scale=0.5]{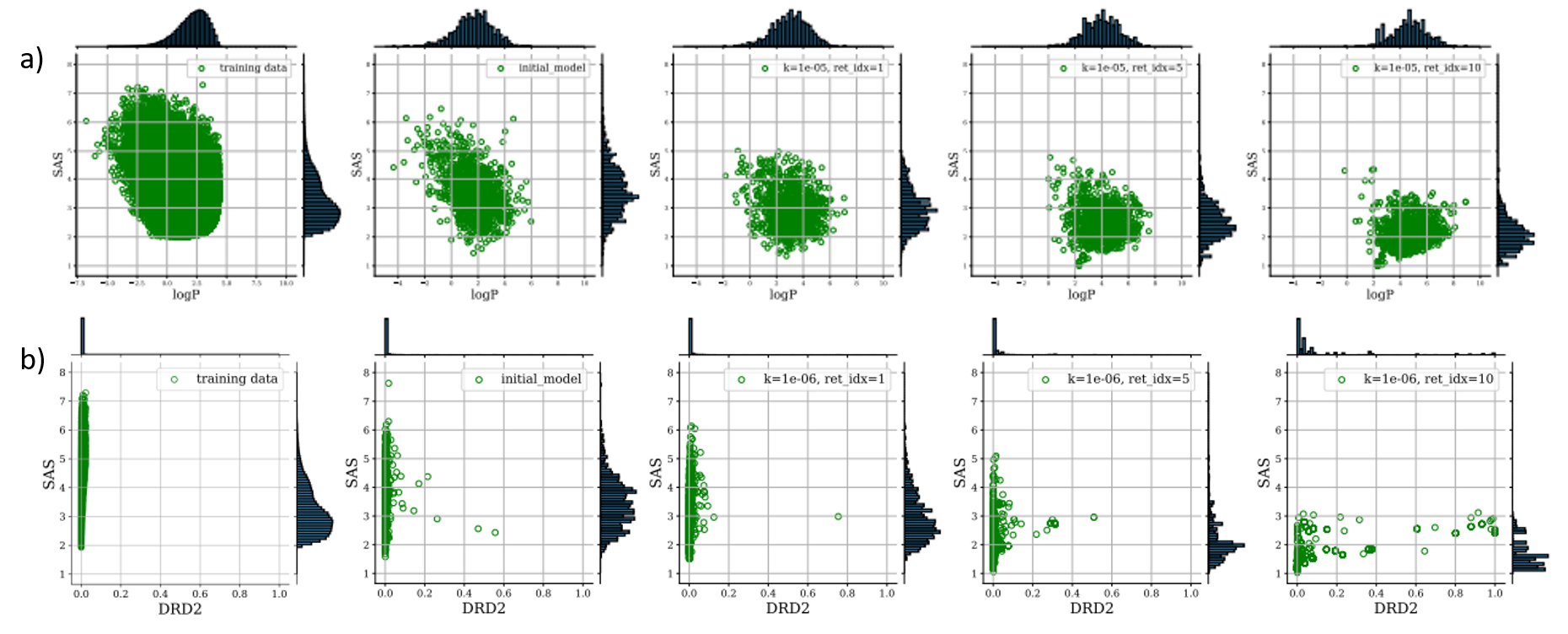}
        \caption{Advancement of the Pareto front from \citeauthor{abeer_multi-objective_2022} using iterative retraining for distribution learning. Both (a) and (b) are from the same optimization task, with each set only showing two objectives for ease of visualization. The first and second columns are the distribution of the training molecules and the first batch of sampled molecules, respectively. The following 3 columns depict molecules sampled from the model after 1, 5, and 10 iterations. Reproduced from \citet{abeer_multi-objective_2022}.
        }
        \label{fig:abeer}
    \end{figure}

    In addition to recurrent neural networks, as in the prior two examples, variational autoencoders and other generative models can be iteratively retrained to simultaneously fulfill multiple property constraints \cite{iovanac_actively_2022}. \citet{abeer_multi-objective_2022} describe one such approach to generate drugs with high predicted binding affinity to the DRD2 receptor, high ClogP, and low synthesizability score using a VAE as the unsupervised generator. After initial training, sampling, and scoring, the best molecules were selected according to their Pareto rank, but some random molecules were also included in the retraining set. Importantly, the authors show a progression of the 2-dimensional Pareto fronts beyond those of the original training set: they identified molecules that are strictly superior to (i.e., that ``dominate'' in a Pareto optimality sense) the best molecules in the training set. Two such plots are shown in Figure~\ref{fig:abeer}. Here, it is clear that this method is capable of increasing the dominated hypervolume and identifying novel molecules that have property values outside of the objective space spanned by the training set.
    
    \subsection{Genetic algorithms}
    
    In contrast to many deep learning architectures, genetic algorithms (GAs) do not rely on a mapping between continuous and discrete spaces. Instead, molecules are iteratively transformed into new ones using \emph{evolutionary operators} like mutations and crossovers. Molecular \emph{mutations} may include the addition or removal of atoms, bonds, or molecular fragments, while molecular \emph{crossover} involves molecular fragment exchange between two parent molecules. GAs begin with a starting population of molecules that are scored by the oracle function(s). Selection criteria are imposed to determine which molecules in the population are chosen as \emph{parents} to be propagated. This selection step is what guides a GA to optimized molecules and, like an acquisition function in BO, determines whether an optimization is a Pareto optimization or not. Evolutionary operators are randomly chosen and applied to the parents, and the population is updated with the resulting molecules.

    Genetic algorithms were the first popularized polymer \cite{venkatasubramanian_computer-aided_1994} and small molecule \cite{sheridan_using_1995} %
    generators. In 1995, \citet{sheridan_using_1995} proposed generating small molecules by iteratively evolving integer sequence representations of molecules. That same year, \citet{weber_optimization_1995} used a GA to find optimal molecules from a synthetically-enumerated library. Since then, GAs have adopted evolutionary operators which function directly on molecular graphs \cite{ pegg_genetic_2001, brown_graph-based_2004, jensen_graph-based_2019} or SMILES strings \cite{nigam_parallel_2022}. Some genetic algorithms even mutate molecules using chemical reaction templates to encourage synthesizability \cite{weber_optimization_1995, durrant_autogrow_2013, daeyaert_pareto_2017}. Multiple objectives can be scalarized during selection to frame a multi-objective GA as a single-objective one \cite{devi_multi-objective_2021, pegg_genetic_2001, jensen_graph-based_2019, herring_evolutionary_2015}.

    As with any generative model, if the selection criteria consider multiple objectives simultaneously without imposing assumptions about relative importance, a GA can advance the population's Pareto front. One such GA was proposed by \citet{brown_graph-based_2004} to generate ``median molecules'', which maximize Tanimoto similarity \cite{bender_molecular_2004} to two different molecules simultaneously. In each iteration, molecules in a population are manipulated with either mutations (add/delete atoms, add/delete bonds) or crossovers (molecular fragment exchange between two parent molecules). Non-dominated sorting, using the two Tanimoto similarities as objectives, determine which molecules are selected for propagation. The critical adaptation for the multi-objective case is the use of Pareto ranking---specifically, NDS---as a selection criterion, instead of using a single property estimate or a scalarization of multiple properties. 
        
\begin{figure}
        \centering
        \includegraphics[scale=1]{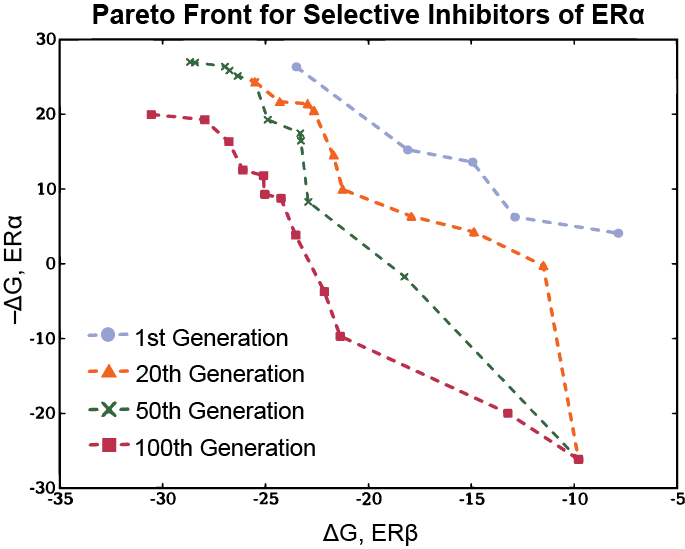}
        \caption{Pareto front for the identification of selective inhibitors. The $\Delta$G values represent docking scores. Note that the Pareto front in this plot is located in the bottom left. The Pareto front is shown after 1, 20, 50, and 100 iterations. It clearly shifts to the bottom left with each iteration. Here, niching is used but elitism is not. Redrawn from \citet{nicolaou_novo_2009}.}
        \label{fig:nicolau-GA}
    \end{figure}

    A comparable multi-objective GA, presented by \citet{nicolaou_novo_2009}, generates ligands with maximized docking scores for a target receptor (Estrogen Receptor $\beta$, or ER$\beta$) and minimized scores for a negative but closely related target (Estrogen Receptor $\alpha$, or ER$\alpha$). As an extension from the prior example, the non-dominated sorting selection criterion was modified to include \emph{niching} and \emph{elitism}. Niching encourages structurally diverse populations by grouping candidates into \emph{niches} based on their structural similarity during selection, and only a set number of molecules may be acquired in each niche. Promoting diversity can be especially beneficial to GA performance, as GAs are constrained by their starting set and set of modification operators \cite{filipic_diversity_2020, zhou_counteracting_2008}. When elitism is imposed, all Pareto-dominant molecules found during prior iterations are appended to the population before selection to prevent good molecules from being ``forgotten.'' The authors report that both elitism and niching improve optimization performance. The depicted progression of the Pareto front is replicated here (Figure~\ref{fig:nicolau-GA}). %
    The notion of optimizing against a negative target can be generalized into a ``selectivity score'' that aggregates affinity to multiple off-target controls \cite{van_der_horst_multi-objective_2012}. 
    
   The effect of diversity-aware acquisition is further explored by \citet{verhellen_graph-based_2022}, wherein the effectiveness of two different multi-objective GAs that promote \emph{Pareto} diversity are compared. Both GAs use non-dominated sorting to select the population members to be propagated as parents of the next generation. %
   The first, NSGA-II \cite{deb_fast_2002}, promotes selection of molecules with a larger distance from other molecules in the objective space and has precedent in application to a synthesizability-constrained molecular GA \cite{daeyaert_pareto_2017}. The second, NSGA-III \cite{deb_evolutionary_2014}, enforces diversity by requiring at least one molecule to be acquired in each of a set of reference regions in the objective space (Figure~\ref{fig:diversity}A). Both genetic algorithms are applied to seven molecular case studies, each with a different set of objectives including affinity to a target, selectivity, and/or molecular weight. Using the dominated hypervolume as an evaluation metric, both multi-objective optimization approaches outperform a weighted-sum scalarization baseline, but there is no clear winner among the two NSGA algorithms. %
   A measure of internal similarity indicates that the structural diversity decreased with each evolutionary iteration. Nonetheless, the selection criteria promoted Pareto diversity, demonstrating that Pareto diversity can be achieved without necessarily requiring molecular, or structural, diversity.

    \subsection{Reinforcement learning} Reinforcement learning (RL)-based generative models are trained to create molecules by learning to maximize a reward function quantifying the desirability of generated molecules. In molecular reinforcement learning, a \emph{policy} determines which molecules are generated and can be iteratively updated to maximize the reward as new molecules are generated and scored. The set of actions or choices available to the policy is denoted the \emph{action space}. The framing of the reward function, analogous to the BO acquisition function and GA selection criteria, determines whether an RL method utilizes Pareto optimization. 
    
    When the learned policy generates molecules by modifying a previous population of molecules, the action space may be comprised of atom- and bond-level graph modifications \cite{ zhou_optimization_2019, leguy_evomol_2020, khemchandani_deepgraphmolgen_2020} or a set of fragment-level graph modifications \cite{erikawa_mermaid_2021}. In a similar manner, graph modifications resulting from chemical reactions can constitute the action space to promote synthesizability \cite{horwood_molecular_2020}. When the policy is a deep learning generator that designs molecules from scratch, any \textit{de novo} generator that decodes latent variables to a molecule, such as SMILES recurrent neural networks, can be considered the policy \cite{olivecrona_molecular_2017, popova_deep_2018, pereira_diversity_2021, neil_exploring_2018, blaschke_reinvent_2020}. Typically, these policies are trained using policy gradient algorithms (e.g., REINFORCE) \cite{williams_simple_1992}.

    Most RL approaches to molecular discovery, and specifically to drug design \cite{tan_reinforcement_2022}, optimize a reward that considers a single property \cite{olivecrona_molecular_2017, popova_deep_2018, pereira_diversity_2021} or a scalarized objective  \cite{neil_exploring_2018, de_cao_molgan_2018, blaschke_application_2018, zhou_optimization_2019, wei_multiple-objective_2019, abeer_multi-objective_2022, leguy_evomol_2020, horwood_molecular_2020, erikawa_mermaid_2021, khemchandani_deepgraphmolgen_2020,  mcnaughton_novo_2022, you_graph_2018, ishitani_molecular_2022, s_v_multi-objective_2022, abbasi_multiobjective_2021}. %
    We are aware of only one molecular RL approach whose reward function directly encourages %
    molecules to be generated along a Pareto front. In  DrugEx v2, presented by \citet{liu_drugex_2021}, RL is used to generate multi-target drug molecules. To promote the discovery of molecules along the Pareto front, NDS is used to calculate the reward. The authors test their algorithm with both this Pareto reward function and a weighted sum reward function. In the weighted-sum benchmark, the weighting factors were set as dynamic parameters which were altered during inference to encourage the model to find solutions at different locations on the Pareto front, analogous to the alternating reward approach to scalarization. For the multi-target discovery case, the fraction of generated molecules deemed desirable (defined as having all properties above some threshold value) was 81\% with the Pareto scheme and 97\% with the weighted sum scheme. The two approaches were only compared in this constraint-style evaluation, not in terms of a Pareto optimization criterion such as hypervolume improvement, so it is not clear if the lackluster performance of the Pareto optimizer is merely due to this misalignment of evaluation criteria. %

    \subsection{Conditional generation} Conditional generators produce molecules that are meant to achieve some set of user-defined properties instead of directly maximizing or minimizing them in an iterative manner. Although our focus in this review is on multi-objective optimization, we feel that discussing the role of conditional generators in MMO is necessary due to their prevalence in the field and the ease of extending from single-objective (single-constraint) conditional generators to multi-objective (multi-constraint) conditional generators. 
    
    Many conditional generators are autoencoders that map molecules to latent embeddings and vice versa. In order to generate molecules with specific properties, the latent variables of these generators can be manipulated during training such that they represent the properties of interest. One such manipulation applied to variational autoencoders is to recenter the prior distribution around the associated molecule's property value $c$ instead of the origin, encouraging the latent distribution to match $\mathcal{N}(c,\sigma^2)$ instead of $\mathcal{N}(0,\sigma^2)$ \cite{richards_conditional_2022, makhzani_adversarial_2016, kang_conditional_2019}. This approach can be expanded to multiple objectives by centering each latent dimension along a different property of interest \cite{richards_conditional_2022}. Then, during inference, sampled latent variables are chosen according to the desired property values with at least partial success.
    
    Autoencoders can also be manipulated for conditional generation by directly feeding the property value(s) of training molecules to the decoder during training \cite{polykovskiy_entangled_2018, simonovsky_graphvae_2018}. As one example, \citet{lim_molecular_2018} use this approach to fulfill certain ``drug-like'' property criteria. During CVAE (conditional VAE) training, a condition vector including molecular weight, ClogP, number of hydrogen bond donors, number of hydrogen acceptors, and topological polar surface area is appended to the latent space during decoding. Then, during generation, a manually specified conditional vector influences the decoder to generate molecules with the stated properties. In all case studies, less than 1\% of generated molecules have properties within 10\% of the values set in the condition vector. Another study using a similar architecture \cite{lee_mgcvae_2022} demonstrates that it is possible for the properties of up to 33\% of generated molecules, ``when rounded up'', to reflect the specified properties. In this case, it appears that this fraction strongly correlates with how many training molecules also fulfilled those constraints.

    Some conditional generators modify existing molecular graphs or scaffolds provided as input instead of generating molecules from scratch. These models are typically trained with matched molecular pairs: pairs of molecules with only one well-defined structural transformation that causes a change in molecular properties \cite{leach_matched_2006, tyrchan_matched_2017}. One such single-objective generative model is intended to ``translate'' molecules that are inactive as DRD2 inhibitors to active inhibitor molecules \cite{jin_learning_2019}, wherein activity is predicted by a trained classifier. The generative model is presumed to learn graphical translations that most contribute to inhibitory strength. 
    This methodology can be extended to the multi-constraint case if improvements in multiple properties are desired \cite{wang_retrieval-based_2022, irwin_chemformer_2022, he_transformer-based_2022}. For example, MolGPT, a conditional generator proposed by \citet{bagal_molgpt_2021}, accepts a scaffold and desired property values. It then outputs a molecule that it believes to fulfill the input constraints. Molecules are completed from scaffolds as SMILES strings, and the model is trained on sets of \{scaffold, molecule, properties\}. %
    The success of MolGPT in meeting target properties relies on having molecules with that property be well-represented in the training set.
    While MolGPT is able to generate molecules conditioned on multiple properties, the authors do not report whether their model is capable of generating molecules with combinations of property values not present in the training set.

     The effectiveness of conditional molecule generators depends not only on their ability to generate valid and unique molecules, but also on the accuracy of the implicit molecule-property model. If this model is inaccurate, the generator will suggest molecules that do not actually exhibit the desired properties. We further emphasize that, in order to identify Pareto-optimal molecules, the model must be able to extrapolate past the training set because, by definition, Pareto-optimal molecules have properties (or combinations of properties) that are not dominated by members of the training set. 
     Therefore, we find it unlikely that these non-iterative conditional generators will succeed in advancing the Pareto front. This is in contrast to iterative optimization methods, wherein the predictive capability of the generators is improved for newly explored regions of chemical space with each iteration. 
     
    Further, the nature of conditional generators requires that a user know what property value ranges are feasible. Based on the discussed and other case studies \cite{kotsias_direct_2020, he_molecular_2021}, conditional generators perform well primarily when attempting to generate novel molecules with property combinations spanned by the training set.  A pIC50-conditioned model  would propose some set of molecules if asked to achieve a pIC50 value of 100, even though such a value is unrealistic. Their behavior in these settings is not well understood, so a user may need to know which property constraints are valid or possible.    
    Due to these concerns, we caution the reader that conditional generators may not be most appropriate for Pareto optimization tasks.

    \subsection{Hybrid approaches} %

    The case studies that we have shared so far fall neatly into our defined categories. However, certain other approaches that combine methods from multiple categories or otherwise deviate from this classification are worth mentioning. 
    
    \citet{grantham_deep_2022} introduce one such hybrid approach, in which latent representations of molecules are mutated with a genetic algorithm and decoded to generate new molecules. %
    A variational autoencoder is first trained to encode molecules into latent vectors. After encoding the starting population, mutations are applied to their corresponding latent vectors, which are then decoded. From this new set of evolved molecules, non-dominated sorting with a crowding distance constraint (specifically, NSGA-II \cite{deb_fast_2002}) is used to select new molecules to use for retraining the autoencoder. %
    The proposed method outperforms two Bayesian optimization baselines in terms of the hypervolume of the final Pareto front when applied to an optimization of ClogP, QED, and synthesizability score. A similar methodology was used to optimize both drug-likeness properties and binding affinity (estimated via docking scores) to carbonic anhydrase IX \cite{mukaidaisi_multi-objective_2022}.  
    
    Iterative retraining has also been used to improve the performance of a conditional generator. In one example, a conditional graph generator is fine-tuned with molecules that are active against both JNK3 and GSK-3$\beta$ \cite{li_multi-objective_2018}. This workflow essentially follows the iterative retraining of distribution learning algorithms, but uses conditional generation to provide an extra bias toward sampling molecules with favorable properties. 
    In a similar manner, reinforcement learning methods can be considered conditional generation if the reward function favors molecules with a target property profile \cite{ domenico_novo_2020, wang_multi-constraint_2021, stahl_deep_2019}. Two such methods \cite{jin_multi-objective_2020, chen_fragment-based_2021} use RL to generate molecules that are predicted to be dual inhibitors of GSK3$\beta$ and JNK3 receptors according to pretrained surrogate models. In the final populations in both studies, 100\% of molecules are active against both inhibitors. However, the dataset used in both studies for training already includes a small fraction of dual inhibitors. Therefore, discovering ``active inhibitors'' in this case is equivalent to discovering the chemical space that is classified as active according to the surrogate models, and this task is easier than extrapolating with a continuous oracle. In general, the reported success of generators conditioned on Boolean values (instead of continuous ones) can be overoptimistic, as the degree of optimization success is harder to quantify with metrics such as the hypervolume.   %

\section{Discussion}

In the description of library-based MMO, we  %
explained that these methods are a natural extension of Bayesian optimization. In contrast, \textit{de novo} methods stray farther from classic BO, although some aspects of BO acquisition functions are present in generative workflows. In particular, NDS is often used as the selection criterion for retraining (distribution learning) or propagation (genetic algorithms). Other conventional BO acquisition functions, such as EHI and PHI, are rarely incorporated into optimization with generative models. These acquisition functions use the uncertainty in surrogate model predictions, which aids in the balance between exploration and exploitation. But most generative optimization architectures score molecules with the ground truth objectives during selection, thus bypassing uncertainty quantification and making EHI and PHI unusable as acquisition functions. An opportunity exists to incorporate Bayesian principles into \textit{de novo} design by including a separate surrogate model that predicts objective function values and can be retrained as new data are acquired to guide selection . These and other adjustments to \textit{de novo} optimization approaches may help bridge the gap between generation and model-guided optimization. 

We have also observed that the performance of Pareto optimization approaches is often evaluated using individual property values or constraints. These metrics, however, reveal little about the \emph{combination of properties} of discovered molecules, which is of foremost interest in MMO. Hypervolume improvement can indicate the shift in the Pareto front, but other qualities of the discovered molecules related to the Pareto front \cite{collette_three_2005, li_quality_2020} can be of equal importance, including the density of the Pareto front or the average Pareto rank of the molecules. In molecular discovery, imperfect property models are often used as oracles. In these cases, it is beneficial to discover a dense Pareto front and many %
close-to-optimal molecules 
\emph{according to QSPR predictions}, even if not all increase the hypervolume. Naturally, some molecules that are predicted to perform well will not validate experimentally, and having a denser population to sample from will increase the probability of finding true hits. For the same reason, promoting structural diversity and not just Pareto diversity is a way to hedge one's bets and avoid the situation where none of the Pareto-optimal molecules validates.

In batched multi-objective optimization, Pareto diversity can be considered during acquisition to promote exploration. In molecular optimization, structural diversity similarly encourages exploration of a wider region of chemical space. Thus, in MMO, both potential measurements of diversity are relevant, and either or both can be used during optimization. At this point, neither diversity metric has been shown to outperform the other in MMO tasks, and the question of how best to incorporate both into acquisition (or whether this actually benefits optimization) remains. At present, diversity-aware acquisition is most commonly incorporated into multi-objective genetic algorithms rather than other generative architectures. Acquisition that promotes diversity may improve performance of generators using reinforcement learning or iterative distribution learning, although this has yet to be demonstrated.  

We have argued that Pareto optimization is a more practical approach to many molecular discovery tasks than scalarization or constrained optimization, but the ability of Pareto optimization to scale to several dimensions must also be addressed. Non-dominated sorting increasingly fails to differentiate the optimality of solutions with more objectives, as more and more points are non-dominated in a higher-dimensional space \cite{maltese_scalability_2018}. The numerical estimation of hypervolume has a computational cost that scales exponentially with the number of objectives, making EHI and PHI acquisition functions also increasingly difficult to use in high dimensions \cite{maltese_scalability_2018}. The increased computational costs associated with fine-tuning many surrogate models and scoring candidates for every objective contribute to scalability issues as well. Considering the challenges faced with Pareto optimization of many (more than three) objectives, scalarizing certain objectives
or converting some to constraints to make the problem solvable may be the most practical approach, especially when some objectives are known to be more important than others. The question of whether Pareto optimization can robustly scale to many objectives is a worthwhile one only if a problem cannot be feasibly reduced. The visualization of the Pareto front is an additional consideration; objective trade-offs are more easily conveyed with a Pareto front of two or three objectives. Ultimately, the optimal formulation of an MMO problem will depend on the use case, and collaboration with subject matter experts can ensure that the problem formulation is feasible but does not impose unrealistic assumptions. 

Beyond these unique challenges posed by multi-objective optimization, many challenges from single-objective optimization remain relevant \cite{bilodeau_generative_2022, renz_failure_2019, meyers_novo_2021}. The first is the need for realistic oracle functions that can be evaluated computationally but meaningfully describe experimental performance; this is closely related to the need for more challenging benchmarks to mimic practical applications.  Optimizing QED, ClogP, or a Boolean output from a classifier are easy tasks and are not good indicators of robustness or generality. Generative models specifically must also prove effective with fewer oracle calls, which is often the bottleneck when molecules must be scored with experiments or high-fidelity simulations \cite{gao_sample_2022}. For experimental applications, the synthesizability of generated molecules is an additional factor that must be considered \cite{gao_synthesizability_2020} and can be cast as a continuous objective or a rigid constraint. Experimental prospective validation is essential to demonstrate the viability of molecular discovery algorithms, though algorithmic advances can be made more rapidly with purely computational studies. %

\section{Conclusion} 

Though many approaches to computer-aided molecular design have been developed with just single-objective optimization in mind, molecular discovery is a multi-objective optimization problem. 
In certain situations, such as optimization from a library (BO-accelerated virtual screening), the extension from single-objective to multi-objective requires only minor modifications, e.g., to the acquisition function and to the number of surrogate models. In contrast, \textit{de novo} design workflows vary more in methodology and are less directly analogous to Bayesian optimization. %
The use of Pareto rank as a reward (for RL) or the use of non-dominated sorting to select sampled molecules to include in subsequent populations (for GAs) or training sets (for iterative distribution learning) replaces greedy acquisition functions. Yet, there is an opportunity to define new generative workflows which more directly incorporate model-guided optimization methods with consideration of model uncertainty. Batching in MMO can encourage chemical space exploration by rewarding structural diversity,  Pareto diversity, or both, but best practices around diversity-aware batching are not well established. %
Emerging workflows will benefit from the adoption of challenging benchmarks and evaluation metrics that measure the dominated hypervolume or Pareto front density. As newly proposed molecular discovery tools increasingly emphasize multi-objective optimization, emerging methods must address the algorithmic complexities introduced by Pareto optimization. %

\section{Acknowledgment}

The authors thank Wenhao Gao, Samuel Goldman, and David Graff for commenting on the manuscript. This work was funded by the DARPA Accelerated Molecular
Discovery program under contract HR00111920025. %

\bibliography{references.bib}

\end{document}